\documentclass[preprint,pra,aps,showpacs,preprintnumbers,amsmath,amssymb,eqsecnum]{revtex4}

\usepackage{graphicx}
\usepackage{dcolumn}
\usepackage{bm}
\usepackage{verbatim}
\usepackage{epsfig}
\usepackage{epstopdf}

\newcommand\beq{\begin{equation}}
\newcommand\eeq{\end{equation}}
\newcommand\bea{\begin{eqnarray}}
\newcommand\eea{\end{eqnarray}}

\def\Q1{{\hat Q(t_1)}}
\def\Q2{{\hat Q(t_2)}}

\def\a{\alpha}

\def\b{\beta}

\def\half{\frac {1} {2}}

\def\x0{{{\bf x}_0}}

\begin{document}


\title{Deccoherent Histories and Measurement of Temporal Correlation Functions for Leggett-Garg Inequalities}

\author{J.J.Halliwell}%
\email{j.halliwell@imperial.ac.uk}

\affiliation{Blackett Laboratory \\ Imperial College \\ London SW7 2BZ \\ UK }



\begin{abstract}
We consider two protocols for the measurement of the temporal correlation functions of a dichotomic variable $Q$ appearing in Leggett-Garg type inequalities. The protocols measure solely whether $Q$ has the same or different sign at the ends of a given time interval, thereby measuring no more than is required for the determination of the correlation function. They are inspired, in part, by a decoherent histories analysis of the two-time histories of $Q$, which yields a number of useful insights, although the protocols are ultimately expressed in macrorealistic form independent of quantum theory. 
The first type involves an ancilla coupled to the system with two sequential CNOT gates, and
the two-time histories of the system (whose probabilities yield the correlation function) are determined in a single final time measurement of the ancilla. It is non-invasive for special choices of initial system states and partially invasive for more general choices. Modified Leggett-Garg type inequalities which accommodate the partial invasiveness are discussed. The quantum picture of the protocol shows that for certain choices of primary system initial state, the final state is unaffected by the two CNOT gate interactions, hence the protocol is undetectable with respect to final system state measurements, although it is still invasive at intermediate times. This invasiveness can be reduced with different choices of ancilla states and the protocol is then similar in flavour to a weak measurement.
The second type of protocol is based on the fact that the behaviour of $Q$ over a time interval can be determined from knowledge of the dynamics together with a measurement of certain initial (or final) data.
Its quantum version corresponds to the known fact that when sets 
of histories are decoherent, their probabilities may be expressed in terms of a record projector, hence the two-time histories in which $Q$ has the same or different sign 
can be determined by a single projective measurement. The resulting protocol resembles the decay-type protocol proposed by Huelga and collaborators (which is non-invasive but requires a stationarity assumption).
\end{abstract}

\pacs{03.65.Ta, 03.65.Ud, 03.65.Yz,  02.50.Cw}







\maketitle

\section{Introduction}

In classical mechanics, when we make sequential measurements on a physical system we can consistently assume that there is an underlying reality, the history of the system, that these measurements reflect, and furthermore, that it is possible in principle, with sufficiently careful measurements, to determine this history without significantly disturbing it.
In quantum mechanics, this is not the case in general. However, we can ask, how far can one go in maintaining a realist picture of the macroscopic realm? The Leggett-Garg (LG) inequalities were proposed to address this question \cite{LG1,L1}.

In the LG framework we consider a dichotomic variable $Q$ which is measured at three or more pairs of times, yielding probabilities of the form $p(s_1,s_2)$ etc, where $s= \pm 1 $ and from these probabilities we construct a correlation function
\beq
C_{12} = \sum_{s_1 s_2} s_1 s_2 p(s_1, s_2).
\label{C12}
\eeq
A realist picture of the macroscopic realm, macrorealism, entails assuming that these observables always take definite values (macrorealism per se, or MRps), that they may be measured without being disturbed (non-invasive measurability, or NIM) and that the state of the system is unaffected by future measurements (arrow of time).
Under these assumptions, it may be shown \cite{LG1,MaTi,ELN} that the correlation functions obey the LG inequalities, which for the case of three pairs of times, may be written
\beq
-1 \le C_{12} + C_{23} + C_{23} \le 1 + 2\ {\rm min} \{ C_{12}, C_{23}, C_{13} \}.
\label{LG}
\eeq
(This concise form was first noted in Ref.\cite{SuZa}).
Numerous experiments have been carried out to test these inequalities and show clear violations, consistent with the predictions of quantum mechanics \cite{Pal,Knee,Rob,Geo,KBLL,KneeMRtest}. A very useful review of both experimental and theoretical aspects of the LG inequalities is Emary et al \cite{ELN}.

What is crucial in these experiments is the way the correlation function is measured. It has to be done in a non-invasive way, otherwise one could argue that the values of the correlation functions were created by an underlying classical model in which the measurements produced disturbing kicks. Indeed, the quantum-mechanical values of the correlation functions can be simulated in precisely this way \cite{Mon,Ye1,Guh,KS}. To achieve this, experiments use, for example, weak measurements \cite{weak,weak2,Wang,Pal,Jor} or ideal negative measurements \cite{LG1,L1,Knee,Rob,KBLL} in which the detector couples to, say, only the $Q=+1$ state at the first time and the absence of detection is taken to imply that the system is in the $Q=-1$ state. Such protocols eliminate or at least restrict alternative hidden variable explanations.

The NIM requirement may also be addressed from the theoretical perspective. Kofler and Brukner proposed that non-invasiveness may be characterized, at least in part, by
the no-signalling in time (NSIT) condition
\beq
\sum_{s_1} p_{12} (s_1, s_2) = p_2 (s_2),
\label{cond}
\eeq
where $p_{12} (s_1, s_2) $ denotes the probability obtained under measurement at both times $t_1$ and $t_2$ and $p_2 (s_2) $ denote the probability obtained under measurement at $t_2$ only, with no earlier measurements \cite{KoBr}. (See also Refs.\cite{MaTi,Cle,Wit} for similar proposals).
This condition is not in general satisfied by the standard quantum-mechanical measurement formula,
\beq
p(s_1, s_2) = {\rm Tr} \left( P_{s_2} (t_2) P_{s_1} (t_1) \rho P_{s_1} (t_1) \right),
\label{2time}
\eeq
where  $\rho$ is the initial state and the projection operators $P_s$ are given by
\beq
P_s = \half \left( 1 + s \hat Q \right),
\eeq
and $P_s (t)= e^{iHt} P_s e^{-iHt} $. However, the NSIT condition is a potentially very useful one and for this reason it was proposed in Ref.\cite{HalQ} to consider the quasi-probability
\beq
q(s_1, s_2) = {\rm Re} {\rm Tr} \left( P_{s_2} (t_2) P_{s_1} (t_1) \rho \right).
\label{2time}
\eeq
This object satisfies NIM
automatically (or more precisely, a generalization of NIM)
and can be measured using weak measurements \cite{weak,weak2,Wang}. It has the same correlation function as Eq.(\ref{2time}) \cite{Fri}.
It can be negative but it was argued in Ref.\cite{HalQ} that the sign of the quasi-probability offers a characterization of the macrorealism broader than that given by the LG inequalities alone. (See also Ref.\cite{Cle}).

Despite these developments and encouraging experimental results, it remains of interest to find new ways of measuring the correlation function in a non-invasive way.
A key issue is that the correlation function is normally determined using sequential measurements and this is the origin of the difficulties of measuring it non-invasively, since the first measurement can affect the value of the second one.
Therefore, it is of interest to find protocols which get away from this feature, for example, by measuring only the correlation function and nothing more.
To this end, note that the correlation function may be written
\beq
C_{12} = p(same) - p(diff),
\eeq
where $ p(same) = p(+,+) + p(-,-) $ and $p(diff) = p(+,-) + p(-,+)$ and note that
\beq
p(same) + p(diff) = 1.
\eeq
Hence we only need to measure $p(same)$  or $p(diff)$ and not the full probability $p(s_1,s_2)$. That is, we need only measure the value of the quantity $ Q(t_1) Q(t_2)$, which takes value $+1$ when $Q$ takes the same sign at both times and $-1$ when $Q$ takes opposite signs. This holds the possibility of being less invasive than a measurement  of the full probability $p(s_1,s_2)$ since less information is determined. However, protocols need to be found to accomplish this and it is not immediately obvious how to do so since $Q(t_1) Q(t_2)$ is a combination of variables at different times. Moreover, in any such protocols, in order to remain within the LG framework, arguments must be given to show that the measurements are equivalent, in a macrorealistic theory, to sequential measurements of the correlation function, Eq.(\ref{C12}). (Inequalities arising from
protocols falling outside this framework in certain respects are most appropriately called Leggett-Garg type  inequalities).

In the quantum-mechanical analysis of this situation, a natural framework to employ is the decoherent histories approach since it is specifically formulated to handle situations in which variables at different times are
involved \cite{GH1,GH2,GH3,Gri,Omn1,Hal2,Hal3,DoK,Ish,IshLin,HalDio,Hal4}. In simple terms, it shows the extent to which the two-time histories of the system can be regarded as entities within themselves, on a par (as closely as possible) with self-adjoint operators at a fixed moment of time. In particular it gives a clear quantum-mechanical framework in which there exists an analogue of the classical object $ Q(t_1) Q(t_2)$. Hence we expect that a quantum histories analysis will give some indications as to how to construct new protocols for measuring the ``same'' or ``different'' histories required for the determination of the correlation function. Any such protocol must be ultimately expressible in purely macrorealistic terms but the quantum formulation is useful for the insights it provides together with the fact that it gives an explicit picture of the dynamics.

The specific aims of this paper are twofold. First, we seek new ways of measuring two-time histories of a dichotomic variable that determine only the correlation function and nothing more. Second, and related, we explore the consequences of the decoherent histories formalism for the two-time histories studied in this context, with a view to shedding light on the measurement of the correlation function.

We will begin in Section 2 by sketching two types of protocols for the measurement of $p(same)$ and $p(diff)$.
The first involves a coupling to an ancilla so arranged that it detects only changes in sign of $Q$ between initial and final times. This protocol is not fully non-invasive in general. However, this can be handled by suitable modifications to the LG inequalities and this is outlined in Appendix A.
The second type of protocol is quite different in character and is based on the idea that $Q(t_1) Q(t_2)$
can in principle be determined, in a macrorealistic theory, from knowledge the initial data at a single time together with some knowledge of the dynamics. It thus requires some assumptions not normally required in the LG framework.
Although both of these protocols were in fact inspired by the quantum-mechanical analysis, and the relationship between the two is clearest in the quantum analysis,
we follow the traditional approach in the LG framework which is to give protocols describable in macrorealistic terms and not dependent on quantum theory.

In Section 3 the decoherent histories analysis is given. (A summary of this formalism for the unfamiliar is given in Appendix B).
We also introduce the class of models we are interested in and describe some of their properties.
The quantum account of the protocol with ancilla is described in Section 4. It follows the classical account of the protocol closely but also suggests new possibilities not apparent in the macrorealistic formulation in terms of NIM and what can be achieved with different choices of initial states.
In Section 5, we show that the decoherent histories formulation offers the possibility of determining the quantum ``history states" directly in a single projective measurement, the quantum analogue of measuring
$Q(t_1) Q(t_2)$ using knowledge of the classical dynamics.
This protocol turns out to bear resemblance to
the approach of Huelga et al \cite{Hue} which regards the question of determining correlation functions as a decay-type problem 
(and also requires a ``stationarity'' condition to relate correlators at different time intervals).
We summarize in Section 6.

This paper has close parallels with another paper by the present author, Ref.\cite{Halvel}, in which a protocol for measuring the correlation function was proposed involving a ``waiting detector'', which interacts with the primary system only at the point in time when $Q(t)$ changes sign. It has a straightforward formulation in macrorealistic terms and is equivalent to the usual correlation function measurement, subject to certain simple assumptions about the underlying dynamics.

On a more general issue, we note that a thorough characterization of macrorealism, critiquing and extending the original one of Leggett and Garg \cite{LG1}, has been given by Maroney and Timpson \cite{MaTi}. They have argued that violations of the LG inequalities rule out only certain types of macrorealistic theories, in particular those of the GRW form \cite{GRW}, but have difficulties with other types such as Bohmian mechanics. The macrorealistic theories we therefore have in mind are of the GRW type.

\section{Protocols for Measuring the Correlation Function}

We seek a method of determining whether the variable $Q$ takes the same or different value at times $t_1$ and $t_2$.
There a number of protocols with which this can be achieved.

\subsection{Protocol with Ancilla}

We suppose that $Q$ starts in some initial state, most generally a statistical mixture of $Q=+1$ and $Q=-1$ states, and
we imagine that $Q$ is coupled to an ancilla at the two times which has two states, $|0\rangle$ and $|1\rangle$ and we suppose that it starts out in the state $|0\rangle$.
(We use a quantum notation but the description is in essence classical).
The coupling is arranged so that at time $t_1$, the ancilla flips states if $Q=+1$ and remains in its original state if $Q=-1$. At time $t_2$, the opposite takes place, i.e. the ancilla flips states if $Q=-1$ and remains in its current state if $Q=+1$. This means that if we find the ancilla in the $|0 \rangle $ state after $t_2$, then $Q$ took values $(-,+)$ or $(+,-)$ at the two times, i.e changed sign, and if we find the ancilla in the $|1\rangle$ state, then $Q$ took values $(-,-)$ or $(+,+)$ at the two times, i.e had the same sign. If many runs are carried out and we note the proportion of times when the two different ancilla states are found, we thus find that the ancilla state probabilities are
\bea
p(0) &=& p(diff)
\\
p(1) &=& p(same)
\eea
Hence the ancilla state detects the ``same'' and ``different'' histories and nothing more.

In general, this procedure is invasive, since an interaction with the ancilla occurs when $Q=+1$ at $t_1$. This could in principle affect the future evolution of $Q$ and hence the state of the ancilla after the second time. There is clearly only one way to avoid this interaction which is to choose an initial state at $t_1$ in which $Q=-1$. This is not a restriction for the simplest spin models since the correlation function is independent of the initial state. To handle the case of initial state $Q=+1$, one could use a different protocol which couples to the opposite signs of $Q$, hence is non-invasive at $t_1$ when $Q=+1$. 
One might then consider handling general initial states by averaging the results of the two protocols (although this may lead to the possibly unsatisfactory situation in which different correlation functions are measured by different means \cite{Ans}). These methods may be sufficient to determine $C_{12}$ and $C_{13}$ non-invasively but there is less control over the value of $Q$ at time $t_2$ so there will be inevitable invasiveness in the measurement of $C_{23}$ since it will have some probability of being in each of the states $Q=+1$ and $Q=-1$. Hence there is a need for modified LG inequalities which allow a certain amount of invasiveness.
This is described briefly in Appendix A (and see also Refs.\cite{Knee,Rob,KBLL}). Loosely speaking, if the invasiveness is sufficiently small, alternative classical explanations of the correlation functions can be outstripped by sufficiently large violations of the LG inequalities.

 One can also address the invasiveness issue by measuring the final state of the primary system after the ancilla state is measured and comparing with its initial state, to determine the extent to which the interaction created any disturbance.
Since there are two interactions, there is in fact the possibility that the primary system state can be disturbed and then returned to its original value by the second interaction. This could be explored experimentally by trying a number of different initial states. We will find that precisely this possibility can occur in the quantum model. However, since there was interaction at the first time, it is difficult to call this non-invasive although it could be called {\it undetectable}.
The possible significance of this will be discussed in the description of the quantum protocol.

Finally, note that due to the close similarity of this protocol to conventional ones measuring the correlation function using sequential measurements, we may assert that the correlation function obtained in this way is the same. The only significant difference here is that the ancilla interaction is specifically arranged to discard unnecessary information. We could, for example, contemplate a more complicated protocol in which a second ancilla is present at time $t_2$ coupled to the $Q=+1$ state (which clearly does not affect the original ancilla interactions). The two ancillas between them perform sequential measurements which determine all four histories of the system. The present protocol then consists of simply discarding the results of the second ancilla.

\subsection{Correlation Function from an Initial Data Measurement}

Since, as noted in the Introduction, sequential measurements are the origin of difficulties with invasiveness, it is of interest to explore protocols which get away from this feature.
The second protocol attempts to do this by considering how the behaviour of $Q(t)$ over the given time interval can be determined from a single measurement of the initial data using knowledge of the underlying dynamics of the system.
It is perhaps best described by an analogy with the arrival time problem \cite{time}. This is the question of determining the probability that a point particle crosses the origin during a given time interval $[t_1,t_2]$ given it's initial state at $t=0$. Classically, this is easily solved in one of two ways for the case of the free particle. The first way is simply to measure the sign of the position at the two times. The second way is to use the classical equations of motion to show that a particle with initial position $x$ and momentum $p$ reaches the origin at arrival time $t= - mx / p$, so the particle crosses if the arrival time lies in the given time interval.
Hence whether the particle crosses or not is determined by a measurement of the specific combination of initial data, $x/p$. Classically, these two different types of measurement are completely equivalent, as long as the sequential measurements of the sign of $x$ are non-invasive. Their equivalence relies on knowledge of the classical dynamics.

We may consider a similar approach applied to the evolution of the dichotomic variable $Q$ (and indeed in some hidden variable models $Q$ is taken to be the sign function of a continuous variable \cite{Mon}, bringing us close to the above analogy). This is difficult to formulate more precisely without detailed knowledge of the dynamics, which is usually not specified in macrorealistic formulations, but it is clearly possible to make some general statements.
We will assume that the underlying dynamics is known to some degree, and in particular that it is possible to determine
the value of $Q$ at time $t_2$ in terms of certain initial data at time $t_1$. In specific quantum models involving spin systems, for example, $\hat Q(t_2)$ is fully determined in terms of $\hat Q(t)$ and its derivative at time $t_1$, as we shall see.
This means that it is possible, at least in principle, to carry out some measurement at a fixed moment of time which will determine whether or not $Q(t)$ has the same or opposite sign at the two times $t_1$, $t_2$. For example, if $Q(t_2)$ is given in terms of $Q(t_1)$ and $\dot Q(t_1)$,
we could contemplate measuring the quantity $Q(t_2) -Q (t_1)$ which is given entirely in terms of $Q(t_1)$ and $ \dot Q(t_1)$. It takes values $ \pm 2$ or zero and yields the correlation function through the relation,
\beq
\langle Q(t_1 )Q(t_2) \rangle = 1 - \frac{1}{2} \langle [Q(t_2) - Q(t _1) ]^2 \rangle
\eeq
Hence there is in principle a single measurable function of the initial data that determines the correlation function. (See related ideas in Ref.\cite{Halvel}). 

In a macrorealistic theory, this type of measurement will give the same correlation function as the one obtained using non-invasive sequential measurements, under the assumption that there exists a classical dynamics relating the function of the initial data to $Q(t_2)$ and $Q(t_1)$. This sort of assumption is clearly stronger than the assumptions usually involved in the LG inequalities, where very little if anything is assumed about the dynamics. However, it does lead to interesting alternative possible measurements of the correlation functions, as we shall see, with the advantage that there is no invasiveness issue since there is only one measurement.
Furthermore,
although rather schematic when stated in a macrorealistic context, this protocol has a very clear implementation
in the decoherent histories analysis of the quantum theory -- under certain 
circumstances there exists a single projective measurement which determines a two-time history, a clear quantum analogue of the above.

\subsection{Relation to the Decay-type Protocol}

We will find in what follows that the decoherent histories implementation of the second protocol resembles
the approach of Huelga et al \cite{Hue}, which formulates the Leggett-Garg situation as a decay-type protocol. In this protocol,
the system is prepared in, say, the $Q=+1$ state at time $t_1$ and $Q$ is then measured at time $t_2$
to see how much has decayed from the $Q=+1$ state, thereby determining $p(same)$ and $p(diff)$.

The similarity may be seen in the following
simple quantum-mechanical sketch.
Suppose we have a two-dimensional Hilbert space with states $ | \pm \rangle$ and take the initial state to be $ |+ \rangle$. Then in the decay-type protocol, we consider
\beq
p(same) = | \langle + | e^{ - i H (t_2-t_1)} | + \rangle |^2,
\eeq
from which the correlation function is obtained.
However, $ e^{ - i H (t_2-t_1)} | + \rangle $ may be expressed as a linear combination of the $|+ \rangle $ and $|- \rangle $ states at the initial time, so $p(same)$ may be regarded as given by a particular type of measurement at the initial time of the $|+ \rangle$ state, namely, the projection onto those parts of the initial state which are still in the $|+\rangle$ state at the later time. Hence given knowledge of the dynamics, the correlation function may be determined in a single projective measurement of the initial state.
The full decoherent histories version, described in Section 5, is however more general than this simple sketch.

Like the above classical protocol, the decay-type protocol clearly satisfies NIM since there are no sequential measurements. However, in order to obtain the correlation for a later time interval $[t_2,t_3]$ some additional assumption is required. Huelga et al made an assumption about the form of the underlying hidden variable theory they call ``stationarity'' \cite{Hue}. Emary et al showed that this could be replaced by a markovianity assumption together with a reasonable assumption of time-translation invariance \cite{ELN}. (However, they note that the status of some of both the assumptions of stationarity and markovianity is somewhat elusive.)

\section{Quantum Histories and Correlation Functions}

\subsection{Decoherent Histories Analysis}

We now consider some properties of the quantum correlation functions and the way they are obtained from the amplitudes for quantum histories.  We follow the decoherent histories approach to quantum mechanics
 \cite{GH1,GH2,GH3,Gri,Omn1,Hal2,Hal3,DoK,Ish,IshLin,HalDio}, which for convenience is summarized in its essential points in Appendix B

The quantum-mechanical probability formula Eq.(\ref{2time}) may also be written, for a pure initial state, as
\beq
p(s_1,s_2) =     \| P_{s_2} e^{ - i H(t_2 -t_1)} P_{s_1} | \psi_{t_1} \rangle \|^2,
\eeq
i.e. as the norm of a  ``history state". From such states we can construct coarse-grained history states describing whether the value of $Q$ is the same or different at the two times, namely,
\bea
|same \rangle &=& \left( P_+ e^{-iH(t_2 -t_1) } P_+ +  P_- e^{-iH(t_2 -t_1) } P_- \right) | \psi_{t_1} \rangle
\nonumber \\
&=&  e^{ - i H t_2 } \left( P_+ (t_2) P_+ (t_1) + P_-(t_2) P_- (t_1) \right) | \psi \rangle
\nonumber \\
&=& \half e^{-iHt_2} \left( 1 + \hat Q (t_2) \hat Q(t_1) \right) | \psi \rangle,
\label{same}
\eea
and also
\bea
|diff \rangle &=& \left( P_- e^{-iH(t_2 -t_1) } P_+ +  P_+ e^{-iH(t_2 -t_1) } P_- \right) | \psi_{t_1} \rangle
\nonumber \\
&=& \half e^{-iHt_2} \left( 1 - \hat Q (t_2) \hat Q(t_1) \right) | \psi \rangle.
\label{diff}
\eea
These states are not normalized and indeed the associated probabilities are
\bea
p(same) &=& \|  | same \rangle \|^2
\nonumber \\
&=& \half \left( 1 + \half \langle \psi | \hat Q (t_1) \hat Q(t_2) + \hat Q (t_2) \hat Q(t_1) | \psi \rangle \right)
\nonumber \\
&=& \half \left( 1 + C_{12} \right),
\eea
where $C_{12}$ is the correlation function, and
\beq
p(diff) = \half ( 1 -  C_{12} ).
\eeq

The two states are not orthogonal but satisfy the condition
\beq
\langle same | diff \rangle  + \langle diff | same \rangle = 0
\label{cons}
\eeq
It is this property, together with the property
\beq
| same \rangle + | diff \rangle = | \psi_{t_2} \rangle.
\eeq
that ensures that the probabilities add up properly
\beq
p(same) + p(diff) = 1
\label{psum}
\eeq
Although coarse-grained history states are defined by summing the amplitudes of the finer-grained history states, the probabilities are in general not related in this way due to interference. An example of this is the fact already noted that  Eq.(\ref{2time}) fails to satisfy the NSIT condition Eq.(\ref{cond}). However, we see here that the ``same" and ``diff" history states do in fact satisfy the correct probability sum rules. This is due to the property Eq.(\ref{cons}), which is a no-interference condition and these properties mean that the histories are consistent but not decoherent, in the language of the decoherent histories approach \cite{GH1,GH2,GH3,Gri,Omn1,Hal2,Hal3,DoK,Ish,IshLin,HalDio}.  This fortuitous property, which does not appear to hold more for more general types of histories, was noted in Ref.\cite{HalDio}.
This property indicates that the histories from which the correlation function is constructed have the same properties as classical histories in terms of their probabilities.

The above is a purely mathematical account of the properties of quantum histories and makes no reference to measurement. Hence to be useful we need to link it to specific measurement procedures. This will be done in the next sections. However, we note that the near-classical properties of the ``same" and ``different" histories suggest that measurement of them may have some simplifying features.

\subsection{A Simple Spin Model}

Consider now specific models to which this framework could be applied. The variable $Q$ is usually taken to be a combination of Pauli matrices so can represent spin systems or the two states of a SQUID, which is what the LG framework was originally designed for.
We focus on a spin system in which $\hat Q = \sigma_z$ and the Hamiltonian is
\beq
H = \half \omega \sigma_x.
\eeq
The dynamics has solution
\beq
\sigma_z (t_2) = \cos  \omega (t_2 - t_1) \  \sigma_z (t_1) + \sin \omega (t_2 -t_1) \ \sigma_y (t_1).
\eeq
The correlation function is given by
\bea
C_{12} &=& \half \langle \psi |  \{ \sigma_z (t_1), \sigma_z (t_2) \} | \psi \rangle
\nonumber \\
&=& \cos \omega (t_2 -t_1),
\eea
and is independent of the initial state, because the anticommutator of any pair of Pauli matrix combinations is proportional to the identity operator.
The commutator at two times is
\beq
[\sigma_z (t_2), \sigma_z (t_1) ] = 2 i \sin \omega (t_2 - t_1)  \ \sigma_x .
\label{comm}
\eeq

More generally, we could take $Q$ to be a general combination of Pauli matrices, $\hat Q = {\bf n} \cdot \sigma $, where ${\bf n}$ is a unit vector. The correlation function is still independent of the state and will have the general form
$C_{12} =  {\bf n} (t_1) \cdot {\bf n} (t_2) $.
Also, the commutator will be
\beq
[ \hat Q (t_1), \hat Q (t_2) ] = 2 i \ {\bf n} (t_1) \times {\bf n} (t_2) \cdot {\sigma},
\label{comm12}
\eeq
which will in general be a different combination of Pauli matrices to those appearing in $H$ and $\hat Q$.

\subsection{Some Useful General Properties}

We now note some general useful and model-independent properties of the operators $\hat Q (t_1)$, $\hat Q(t_2)$. Define the hermitian operators
\bea
\hat C &=& \half \{  \hat Q(t_1), \hat Q(t_2) \},
\\
\hat D &=& \frac {i} {2} [ \hat Q(t_2), \hat Q(t_1)].
\label{D}
\eea
(The operator $\hat C$ is of course the correlation function multiplied by the identity operator in the simple spin case and $\hat D$ is given by Eq.(\ref{comm}).
Then some useful properties follow solely from the fact that $\hat Q^2 = 1$, namely
\bea
[ \hat Q(t_1), \hat C ] =&0&= [ \hat Q(t_2), \hat C],
\\
\{ \hat Q(t_1), \hat D \} =&0&= \{ \hat Q(t_2), \hat D \}.
\eea
It follows from these two relations that
\beq
[ \hat C, \hat D] = 0.
\eeq
This means that $\hat C$ and $\hat D$ have common eigenstates. Also it is easily shown that
\beq
\hat C^2 + \hat D^2 = 1.
\label{squares}
\eeq
We will exploit these properties below.

These properties allow us to say something about the state-dependence of the correlation function in the general case without appealing to the Pauli matrix form.
Since $\hat C$ commutes with $\hat Q(t_1)$ we may work in a basis consisting of eigenstates of $\hat Q(t_1)$, defined by
\beq
\hat Q (t_1)  | \pm, \nu_\pm \rangle = \pm | \pm, \nu_\pm \rangle,
\eeq
where $\nu_\pm$ is the degeneracy. $\hat C$ is clearly diagonal in these states.
Noting that $\hat D$ anticommutes with $ \hat Q (t_1)$, we see that $\hat D$ maps the $+$ states into the $-$ states and vice versa. If the $+$ and $-$ states have the same degeneracy, we set $\nu_\pm = \nu$ and it
is convenient to choose these states so that
\beq
| -, \nu \rangle = \frac {1} {  \langle \hat D^2 \rangle^\half} \hat D | +, \nu \rangle,
\label{+-states}
\eeq
where the average in the normalization factor is taken in the $+$ state.  However, the degeneracies may be different, for example, if we have a three dimensional Hilbert space with
\beq
\hat Q = | 1 \rangle \langle 1 | -  | 2 \rangle \langle 2 | - | 3 \rangle \langle 3 |.
\eeq
In that case, and if, say the $+$ subspace has the smaller degeneracy, we can simply use Eq.(\ref{+-states}) to define class of $-1$ eigenvalue states and then work with this less than complete set. It is fine to do that since LG tests do not require us to work with a complete set of initial states (and indeed different values of the correlation function are usually obtained by adjusting the time intervals).

It follows from the above that
\beq
\langle -, \nu | \hat C | - ,\nu \rangle = \frac {1} {  \langle \hat D^2 \rangle}
\langle +, \nu |  \hat D \hat C \hat D | + \nu \rangle.
\eeq
We now note that $\hat D$ commutes with $\hat C$ and also
\beq
\hat D^2  | + \nu \rangle =   {  \langle \hat D^2 \rangle} | +, \nu \rangle,
\eeq
since, by Eq.(\ref{squares}), an eigenstate of $\hat C^2$ is also an eigenstate of $\hat D^2$. It therefore follows
that
\beq
\langle -, \nu | \hat C | -, \nu \rangle =\langle +, \nu | \hat C | +, \nu \rangle.
\eeq
The consequence of these features for the correlation function is as follows. Suppose we compute the correlation function in a superposition state,
\beq
| \psi \rangle =   a_1  |+, \nu \rangle  +a_2   |-, \nu \rangle,
\label{a1a2}
\eeq
where $|a_1|^2 + |a_2|^2 = 1$. We have
\bea
C_{12} &=& \langle \psi | \hat C | \psi \rangle
\nonumber \\
&=& |a_1|^2   \langle +, \nu | \hat C | +, \nu \rangle + |a_2|^2 \langle -, \nu | \hat C | -, \nu \rangle
\nonumber \\
&=& \langle +, \nu | \hat C | +, \nu \rangle.
\eea
Hence the correlation function has some dependence on the state but if we define a subset of states of the form Eq.(\ref{a1a2}), with $ | \pm, \nu \rangle$ as defined above, the correlation function is independent of the states within that subset.  Since $\hat C$ also commutes with $\hat Q (t_2)$, an identical argument holds in terms of eigenstates of $ \hat Q(t_2)$.

\section{The Quantum Protocol with Ancilla}

We now give a quantum account of the protocol with ancilla described in Section 2 making use of the quantum histories formalism developed in the previous section. This is in effect a method to measure the history states, $|same\rangle$, $|diff \rangle$.

We suppose the system can be interacted with an ancilla consisting of a two state system which has orthogonal states $|0\rangle$ or $|1\rangle$ and we use CNOT gates to flip between these two states depending on the value of $Q$ at each time. (An ancilla was used in this context in Ref.\cite{PaMa} and a CNOT gate in Ref.\cite{Knee}, but in a different way.) We consider two cases. In the first, the ancilla initial state is $|0\rangle$ and in the second it is more general.

\subsection{Simple Initial Ancilla State}

We suppose the combined system starts in the initial state $|\psi \rangle \otimes |0 \rangle$ at $t=0$ and evolves unitarily to time $t_1$, at which point it meets a CNOT gate which flips the auxiliary system if $Q=+1$ and leaves it alone if $Q=-1$.
The state after this interaction is thus
\beq
|\Psi_1 \rangle =  P_+ | \psi_{t_1} \rangle \otimes |1 \rangle + P_- |\psi_{t_1} \rangle \otimes | 0 \rangle.
\label{Psi1}
\eeq
The combined system is then evolved unitarily to time $t_2$ at which point it meets another CNOT gate with the opposite effect: it leaves the auxiliary system alone if $Q=+1$ and flips it if $Q=-1$. The state afterwards is therefore,
\bea
| \Psi_2 \rangle =
P_- e^{-iH(t_2 -t_1) } P_+ | \psi_{t_1} \rangle \otimes |0\rangle
+ P_+ e^{-iH(t_2 -t_1) } P_+ | \psi_{t_1} \rangle \otimes |1\rangle \nonumber \\
+ P_- e^{-iH(t_2 -t_1) } P_- | \psi_{t_1} \rangle \otimes |1\rangle
+ P_+ e^{-iH(t_2 -t_1) } P_- | \psi_{t_1} \rangle \otimes |0\rangle.
\eea
This is conveniently rewritten in terms of the history states introduced in the previous section,
\beq
| \Psi_2 \rangle = | same \rangle \otimes |1 \rangle + | diff \rangle \otimes |0\rangle.
\label{Psi2}
\eeq
Eq.(\ref{Psi2}) means that the probabilities for the ``same" and ``diff" histories can be obtained by measuring the ancilla at a single time, without involving sequential measurements, as described in the classical version.

The system and ancilla have interacted at two times, so in general there is a measurement disturbance at $t_1$ (except in the case in which the state at time $t_1$ is an eigenstate of $\hat Q$, consistent with the classical account).
However, the disturbance to the system at the final time can be avoided by a careful choice of initial state,
which we have some freedom to choose if the correlation function has no (or limited) dependence on it.
To this end, note that we may write
\bea
| same \rangle &=&  \half  e^{-iHt_2} \left( 1 + \half \{ \hat Q (t_2), \hat Q(t_1)\} + \half [ \hat Q(t_2), \hat Q(t_1) ]
\right) | \psi \rangle
\nonumber \\
&=& \half \left( 1 + \hat C - i \hat D \right) | \psi \rangle.
\label{CD}
\eea
Since as shown, $\hat C$ and $\hat D$ commute, we may {\it choose} the initial state to be an eigenstate of both operators. For the simple spin case, $\hat C $ is just $C_{12}$ times the identity and $\hat D$ is proportional to $\sigma_x$, so we would then take the initial state to be a $\sigma_x$ eigenstate. We thus obtain
\beq
| same \rangle = \half \left( 1 + \langle \hat C \rangle -  i \langle \hat D \rangle  \right)
| \psi _{t_2} \rangle,
\eeq
and we note that $ \langle \hat C \rangle = C_{12} $.
The $ | same \rangle $ state is then simply the unitarily evolved initial state multiplied by a c-number, with a similar result for $| diff \rangle$. We thus obtain the striking result that the final state of the combined system Eq.(\ref{Psi2}) has the factored form
\beq
| \Psi_2 \rangle =  |\psi_{t_2} \rangle  \otimes \left(  a |1 \rangle + b | 0 \rangle \right),
\label{factored}
\eeq
where
\bea
a &=&  \half \left( 1 + C_{12}  -  i \langle \hat D \rangle \right),
\\
b&=&   \half \left( 1 - C_{12}  +  i \langle \hat D \rangle \right).
\eea
In particular, the system state is exactly the same as it would have been in the absence of interactions.
Using the property Eq.(\ref{squares}), together with the fact that $\langle \hat D^2 \rangle = \langle \hat D \rangle^2$ (since the state is an eigenstate of $\hat D$), and likewise for $\hat C$,
we find
\bea
p(1) = &=& |a|^2 = \half ( 1 + C_{12} ) = p(same),
\nonumber \\
p(0) = &=& |b|^2 = \half ( 1 - C_{12}) = p(diff),
\eea
as expected.

Since $\hat C$ and $\hat D$ will generally have a number of eigenstates, we could construct a family of density matrices diagonal in those states. Using this density matrix as the system initial state will yield essentially the same results as the pure state case given above.

It may seem surprising that in the factored state Eq.(\ref{factored}), it is possible to extract the correlation function without any disturbance to the final system state. However, this is a question of the degree of dependence of the correlation function on the system state. For the simplest spin systems, the correlation function is independent of the state. For more general ones, note that the result Eq.(\ref{factored}) holds for only a special class of initial states, namely eigenstates of $\hat D$. These can be expressed in the form Eq.(\ref{a1a2}) for suitable coefficients $a_1$, $a_2$ but the correlation function is then independent of those coefficients, hence it is independent of the state within a limited class. For more general systems and initial states there will be state-dependence in the correlation function but the final combined system state will not be of the factored form Eq.(\ref{factored}).

After the interaction at the first time, the system state is disturbed, as one would expect -- tracing out the auxiliary system reveals a mixed system state whereas the initial system state is pure. However, with the above choice of initial state, the second interaction causes the state to return to factored form, hence the overall effect after time $t_2$ is that the system state is exactly the same as it would have been if there were no interactions.
The measurement interaction is undetectable in terms of system measurements on the final state, but
the intermediate invasiveness may allow alternative classical explanations as discussed in Section 2.

The quantum account of the protocol yields a new feature not present in the classical account. In the discussion of alternative classical explanations in Section 2, it was assumed that the ancilla gave nothing more than classical kicks to the system and the question is then whether a system described by dichotomic variable $Q$ with classical kicks could replicate the quantum correlation functions, as in Ref.\cite{Mon}. However, the quantum case involves entanglement
between the system and ancilla and the way in which the system is affected at each interaction is more subtle -- the system wave function is not collapsed by the interaction. In addition, there is the intriguing property that with the special initial state above the combined system goes from factored to entangled and back to factored again. It is therefore of interest to regard the system-ancilla combination as a single quantum system with Hilbert space of dimension four or more and then ask whether the properties of this composite system can be classically replicated. One would expect that this is not possible in general \cite{KS}. (See also the related question of whether the three box problem has a classical explanation \cite{Mar}). That is, although the protocol appears to be invasive when formulated in macrorealistic terms, it may resist classical replication due to the dimension of the composite Hilbert space and the specific form of the entanglement that takes place between system and ancilla. However, such an explanation departs from the standard LG framework so will not be pursued here.

Finally, it is of interest to comment on the role of the no-interference condition Eq.(\ref{cons}) in this protocol, or what is equivalent, the probability sum rule Eq.(\ref{psum}), which as stated earlier does not hold in general for histories involving combinations of amplitudes at different times. The measurements on the ancilla must clearly satisfy
\beq
p(1) + p(0) = 1,
\eeq
for any state and with the state Eq.(\ref{Psi2}) this then implies Eq.(\ref{psum}). This means that if the no-interference condition Eq.(\ref{cons}) did no hold it would not be possible to obtain an entangled state of the form Eq.(\ref{Psi2}). Hence the protocol relies on the near-classical property of histories, Eq.(\ref{cons}).

\subsection{More General Initial Ancilla States}

It is of interest to ask whether the protocol with ancilla may be modified in some way so as to minimize the disturbance after the first measurement.
To this end, we note that we have the possibility to explore different initial states for the ancilla.
We therefore consider a modified protocol in which the ancilla starts out in the (normalized) state $ \alpha |0\rangle + \beta |1 \rangle$.
The state after the first measurement, Eq.(\ref{Psi1}), is then
\beq
|\Psi_1 \rangle =  P_+ | \psi_{t_1} \rangle \otimes \left( \alpha |1 \rangle + \beta |0 \rangle \right)
+ P_- |\psi_{t_1} \rangle \otimes \left( \alpha | 0 \rangle + \beta | 1 \rangle  \right).
\eeq
Noting that $P_{\pm} = \half (1\pm \hat Q )$, this may be written
\beq
| \Psi_1 \rangle = \half (\alpha + \beta) | \psi_{t_1} \rangle \otimes ( |1\rangle + |0 \rangle )
+ \half (\alpha - \beta) \hat Q | \psi_{t_1} \rangle \otimes ( |1\rangle - |0 \rangle ).
\eeq
We thus see that if $\alpha$ is very close to $\beta$, the change in both system and ancilla state is very small at the first time, and zero for $\alpha = \beta$.

Evolving to the second time, we find that
\bea
|\Psi_2 \rangle &=& | same \rangle \otimes  \left( \alpha |1 \rangle + \beta |0 \rangle \right)
+ | diff \rangle \otimes (  \alpha |0\rangle + \beta |1 \rangle)
\nonumber \\
&=& ( \alpha |same \rangle + \beta |diff \rangle ) \otimes |1 \rangle
+ ( \beta | same \rangle + \alpha | diff \rangle ) \otimes |0\rangle.
\eea
The two history states, ``same" and ``diff" are now correlated with two ancilla states, but the ancilla states are generally not orthogonal and in particular will not be in this case since we require $\alpha$ and $\beta$ to be close. Interestingly, this does not matter! For suppose we look for the probability of finding the ancilla in the $|1 \rangle $ state, we find
\beq
p(1) = | \alpha |^2 \langle same | same \rangle + | \beta |^2 \langle diff | diff \rangle
+ \alpha^* \beta \langle same | diff \rangle + \beta^* \alpha \langle diff | same \rangle.
\eeq
We may, for convenience choose $\alpha$ and $\beta$ to have the same phase, in which case the last two terms cancel due to the no-interference property Eq.(\ref{cons}). We thus obtain
\bea
p(1) &=& \half \left(  1+ (| \alpha|^2 - | \beta|^2)  C_{12} \right),
\\
p(0) &=& \half \left(  1- (| \alpha|^2 - | \beta|^2)  C_{12} \right).
\eea
Hence if we know the initial state, we can read off the value of the correlation function from either $p(1)$ or $p(0)$.

We thus see that choosing an initial ancilla state with $\alpha$ close to $\beta$, the disturbance due to the first measurement can be made arbitrarily small. By choosing the system initial state in the way described in Section 3, we can ensure that the system state after the second measurement is exactly as if no measurement was made.
This protocol is therefore undetectable with respect to final system measurements and also satisfies the standard version of NIM, to a controllable degree, dependent on the precision to within which one can measure $p(1)$ and $p(0)$.

This protocol is similar in flavour to weak measurements,
in that the desired quantity is found by examining the small bias around the zero disturbance result $\alpha = \beta$. The measurements here are not weak -- the disturbance is rendered small by carefully choosing the initial state so that the CNOT gates do not disturb them very much. However, this protocol has no obvious expression in macrorealistic terms. Furthermore, its similarity with weak measurements may mean it is susceptible to the same criticisms \cite{ELN,deco}. For example, the fact that the measurement disturbance is typically of the same order of magnitude as the small bias from which the correlation function is extracted raises questions as to whether the measurement is really non-invasive in any sense.
However, in this case, the measurement disturbance at time $t_1$ is zero if the system is in the $Q=-1$ state at that time, as discussed, hence the disturbance can be kept smaller than the effect we are measuring by restricting to system states in that neighbourhood. (This is of course at the expense of losing the undetectability property described above but the non-invasiveness is the most important property to maintain).
See also Ref.\cite{Halvel} for another example along these lines in which a weak measurement is arguable non-invasive.

\section{Decoherent Histories and Single Time Measurements}

We saw in Section 4, through Eq.(\ref{Psi2}), how the history states could be determined through measurements of the ancilla states. The history states themselves cannot be determined directly through projective measurement since they are not orthogonal, as noted. Indeed, we have
\bea
\langle same | diff \rangle &=&  \frac {1} {4} \langle \psi |  [ \hat Q (t_1), \hat Q(t_2) ] | \psi \rangle
\nonumber \\
&=& \frac {i} {2} \langle \hat D \rangle
\eea
Hence they are orthogonal for initial states in which $ \langle \hat D \rangle = 0$, which is $ \langle \sigma_x \rangle = 0$ for the simple spin model.

When the history states are orthogonal,  the decoherent histories formalism offers another way of relating the description of histories in terms of a string of projection operators to a single measurement. The point is that when orthogonal the history states may be associated with a projector, a so-called ``record'' projector $R$, which is perfectly correlated with the history states. This important feature of the decoherent histories approach is described in detail in Appendix B.

In the simple case considered here, this projection may be taken to be
\beq
R_{same}    = e^{  i H t_2} \frac{  | same \rangle \langle same | } { p(same)} e^{ - i H t_2}
\eeq
(The extra unitary time evolution factors in comparison to Eq.(\ref{rec}) are due to the fact that the history states Eq.(\ref{hist}) differ in their definition by similar factors from the history states Eqs.(\ref{same}), (\ref{diff}).)
The probabilities for histories may then be expressed in the standard form, as the average of this projector:
\beq
p(same) = \langle \psi | R_{same}  | \psi \rangle.
\label{probR}
\eeq
In other words, in order to determine the probability for a two-time history, characterized by pairs of non-commuting projection operators, we need only make the single time measurement defined by the projection operator $R_{same}$.
This can be thought of as a measurement of the record of the history stored somewhere in the system, as discussed in Appendix B. But it can also be regarded, by unitary shifts in time,
as a projection on the initial state, i.e. restriction to those parts of the initial state which lead to $Q$ keeping the same sign, hence is an expression of the second protocol outlined in Section 2B.

In general, the projector $R_{same}$ could be quiet a complicated function of $\hat Q$ at different times and may not have a simple expression in terms of quantities that are easily measured. However, it can clearly be determined in simpler form in specific models. Indeed, for the simple spin case,
from Eq.(\ref{CD}),  and using  $\hat C = C_{12}$ (times the identity) and $\hat D$ given by Eqs.(\ref{D}), (\ref{comm}), we find
\bea
| same \rangle &=& \half e^{- i H t_2}  \left( 1 + \cos \omega (t_2 - t_1) + i \sin \omega (t_2 - t_1) \sigma_x  \right)
| \psi \rangle,
\nonumber \\
&=& \half e^{ - i H t_2} \left( 1 + e^{ 2 i H (t_2 - t_1) } \right) | \psi \rangle,
\nonumber \\
&=& \cos ( H(t_2 - t_1) ) | \psi_{t_1} \rangle,
\nonumber \\
&=& p(same)^{\half}  | \psi_{t_1} \rangle.
\eea
Hence we get the suprisingly simple result that it is just the initial state at $t_1$ times a numerical factor. This now means that
\beq
R_{same} = e^{  i H(t_2 - t_1) } | \psi \rangle \langle \psi | e^{-  i H (t_2 - t_1) },
\eeq
and therefore the probability Eq.(\ref{probR}) is
\beq
p(same) = | \langle \psi_{t_2}  | \psi_{t_1}  \rangle |^2.
\label{decay}
\eeq
One can easily confirm that this formula gives the expected correlation function.

In measurement terms, this means that for all initial states satisfying $ \langle \sigma_x \rangle = 0$, $p(same)$ is obtained by simply preparing the system at $t_1$ and then counting the proportion of runs in which it is found in the same state at $t_2$. This is clearly very similar to the decay-type protocol described in Section 2C, but with here the difference that
the initial and final states do not have to be $\sigma_z$ eigenstates but can be any rotations of such eigenstates within the $yz$ plane.

Note that the decoherence condition $ \langle \hat D \rangle =0$, which becomes $\langle H \rangle = 0$ in the simpler models, is simply a restriction to initial states which undergo a transition under $H$. Only for such states does the formula Eq.(\ref{decay}) give the correlation function. If the initial state was an eigenstates of $H$, and therefore not satisfy the decoherence condition, Eq.(\ref{decay}) would yield a constant.

The above construction of the projective measurement easily generalizes to the case of any Hamiltonian with $H^2$ proportional to the identity (with initial states such that $ \langle H \rangle = 0$).
For more general cases, the projector always exists and can easily be calculated in a specific model, but the probability does not obviously have the simple form Eq.(\ref{decay}).

\section{Summary}

This purpose of this paper was to explore two different but related themes in connection with the correlation functions used in the Leggett-Garg inequalities and their generalizations. The first theme was to find new ways of measuring two-time histories of a dichotomic variable that determine only the correlation function and nothing more. The second was to explore the consequences of the quantum histories formalism for the two-time histories studied in this context.
It is of interest to explore both of these themes since traditional measurements of the correlation functions are experimentally demanding so alternative protocols are clearly potentially useful. Furthermore, alternative protocols could give new perspectives on non-invasiveness.

In Section 2, we proposed, in macrorealistic terms, a protocol involving an ancilla for measuring probabilities for the ``same" and ``different'' histories of the dichotomic variable $Q$ from the which the correlation function is obtained. It measures only whether $Q$ is the same or different sign at the two times and nothing more, so is less invasive than traditional protocols. It is, however, partially invasive in that there is an interaction with one of the possible values of $Q$ at each time.
This can be made zero with a special choice of initial state in the measurement of $C_{12}$ and $C_{13}$ and more general states can be incorporated by combining two different protocols for the ancilla, but there is inescapable invasiveness in the measurement of $C_{23}$ since there is less control over the value of $Q$ at time $t_2$. Modified LG inequalities taking into account sources of invasiveness were discussed in Appendix A, and they indicate that 
alternative classical explanations for the quantum correlation functions can be outstripped by
sufficiently large violations of the inequalities.
We also proposed a second protocol, in which it was argued in general terms that the variable $Q(t_1) Q(t_2)$ could be determined by a suitably chosen measurement of initial (or final) data, subject to suifficient knowledge of the classical dynamics. This could be expressed only in a schematic way at the macrorealistic level, but has a more concrete realization in the quantum case.

In Section 3, we gave a decoherent histories analysis of the two-time histories and constructed the quantum states $|same\rangle$ and $|diff \rangle$ describing the histories in which $Q$ changes sign or not. Some properties of the correlation functions were also described, for reasonably general systems.

The decoherent histories analysis does not correspond directly to measurement of the system, so protocols need to be specified to show how the history states are determined in a specific measurement context. Two different methods were proposed to accomplish this. The first was the quantum version of the protocol with ancilla.
The quantum case follows the classical one closely. However, it led to new possibilities. In particular, in Section 4(A), we saw that it is possible to choose initial system states so that the final system state, after two interactions, carries no trace of the interactions, and thus the protocol is undetectable in terms of future system measurements. However, there is still intermediate interaction so it is not non-invasive in the conventional definition. Nevertheless, it is still an interesting question to determine whether the quantum properties of the system-ancilla composite system can be replicated in a classical way, although we did not do so here. General results suggest it is not possible but this is a matter of building specific models. Furthermore, this falls outside the usual scope of the LG framework, in which everything is formulated in macrorealistic terms independent of quantum theory. Still, even if outside this framework, it remains of interest to determine whether a given quantum situation has any kind of classical model (as studied previously in the three box problem, for example \cite{Mar}).

Similar remarks apply to the exploration of different ancilla states in Section  4(B). There we found the protocol of Section 4(A) could be made to look minimally invasive at the first time, in a way akin to weak measurements. However, the formulation seems to be purely quantum with no obvious macrorealistic account.

The second method of measuring the histories states, described in Section 5, is inspired directly by a key property of the decoherent histories approach to quantum mechanics. This is the feature that quantum histories, normally represented by a string of non-commuting projection operators, can be completely represented by a single projection operator at a fixed time if the histories are decoherent (i.e. the histories states are orthogonal). This means that there is a single measurement that collapses onto the $|same \rangle$ state. The form of this projective measurement was determined explicitly for a simple (but relevant) class of systems and found to be essentially a projection onto the initial state evolved to the final time. This is therefore the quantum account of the second protocol described in Section 2B. 
Moreover the final formula for $p(same)$ in simple examples coincides with
the closely related decay-type protocol, 
but generalized to a wider class of initial states.

\section{Acknowledgements}

I am very grateful to George Knee for useful comments about this work and also to James Yearsley for numerous conversations about the LG inequalities over a long period of time. I would also like to thank three anonymous referees for their constructive comments.

\appendix

\section{Leggett-Garg Inequalities with Partial Invasiveness}

The ancilla-based protocol described in Section 2 is, as noted there, only partially invasive in general, except for special initial states. One is therefore faced with the question of to what degree this partial invasiveness
could explain the LG inequality violations observed in experiments. Differently put, how large does the inequality violation have to be in order to outstrip classical explanations? One can also ask about the degree to which clumsy ancilla preparation may affect the LG inequalities. These sorts of issues have been addressed in Refs.\cite{Knee,Rob,KBLL}. Here, we focus on the partial invasiveness arising in the ancilla protocol in Section 2.

We follow the general mathematical framework known as the ``contextuality by default'' approach of
Dzhafarov and Kujala \cite{DzKu}. (See also the discussion of this approach by Bacciagalupi \cite{Bac}). In this approach, we can measure a variable $A$ on its own or measure it in the context in which another variable $B$ is also measured. If the measurement of $B$ changes the probability distribution of $A$, then $A$ is not the same random variable in the two contexts. It then makes sense to denote the variable in the second context by $A^B$ and regard $A$ and $A^B$ as independent random variables described by an underlying probability distribution.

Applied to the LG system, we consider $Q(t)$ at the three times, which we denote for convenience  $Q_1, Q_2, Q_3$, and given that we do three pairwise measurements, we note that each variable can be measured in two different contexts. So $Q_2$ can be measured as part of the pair of measurements at $t_1$ and $t_2$, or part of the pair of measurements at $t_2$ and $t_3$. We denote the values of $Q_2$ obtained in these two different contexts by $Q_2^{12}$ and $Q_2^{23}$. Clearly in the first case the earlier measurement at $t_1$ can affect the value of $Q_2$ but in the second case, there is a later measurement, which cannot affect $Q_2$ (assuming the arrow of time). Similarly, $Q_3$ can be measured in the two contexts in which measurements were made at $t_1$ and $t_3$ or at $t_2$ and $t_3$, and we denote the results $Q_3^{13}$ and $Q_3^{23}$.
Finally $Q_1$ can be measured in the two possible contexts in which later measurements were made at $t_2$ or at $t_3$, but, assuming the arrow of time, $Q_1$ will be the same in both cases.

Dzhafarov and Kujala showed that in the presence of such invasive effects,
the LG inequalities take the modified form,
\beq
-1 - 2 \Delta_0  \le C_{12} + C_{23} + C_{23} \le 1 + 2 \Delta_0 + 2\ {\rm min} \{ C_{12}, C_{23}, C_{13} \},
\label{LGmod}
\eeq
where 
\beq
\Delta_0 = \half \left(    | \langle Q_2^{12} \rangle -  \langle Q_2^{23} \rangle | 
+ | \langle Q_3^{13} \rangle -  \langle Q_3^{23} \rangle | 
\right).
\label{Delta}
\eeq
These inequalities may be used to distinguish between the genuine contextuality of the type exhibited by quantum mechanics from the ``cross-influences'' coming from invasive measurements.
If violated, it means that the correlation functions cannot be explained by a classical theory with invasive kicks.

These modified inequalities can be applied to the ancilla protocol in Section 2. In the simplest case in which the initial state is $Q=-1$, the measurement is non-invasive at $t_1$, which means that
$Q_2^{12} = Q_2^{23} $, so the first term in $\Delta_0$ drops out. The second term is then a measure of the invasiveness at $t_2$. More generally, we can use this framework to consider arbitrary initial states which would then have invasiveness at $t_1$ also. We expect that if the invasiveness is sufficiently small, violations of these modified inequalities will be possible.

Note that to test these inequalities experimentally it is necessary to determine the values of $\langle Q \rangle $ at the initial and final times in each pair, in addition to the correlation function. However, this is readily achieved by adding a second ancilla at the second time, as outlined at the end of Section 2(A).

For more detailed calculations with these modified inequalities see Ref.\cite{Bac}.
These issues will be taken up in greater detail in a future publication.

\section{The Decoherent Histories Approach to Quantum Mechanics}

Here the key aspects of the decoherent histories approach are summarized \cite{GH1,GH2,GH3,Gri,Omn1,Hal2,Hal3,DoK,Ish,IshLin,HalDio,Hal4}. This is described in many places but the brief account here follows Ref.\cite{HalDio} quite closely. The goal of this approach is to assign probabilities to the histories of a closed quantum system, without appealing to measurements undertaken by an external classical domain,
and thereby determine the situations in which statements about quantum systems may be manipulated according to classical logic.

A history is a sequence of alternatives, for example the values of spin vairables in a given direction,
at as sequence of times. In quantum theory, alternatives at each moment of time are
represented by a set of projection operators $\{ P_a \}$,
satisfying the conditions
\bea
\sum_a P_a &=& 1,
\\
P_a P_b &=& \delta_{ab} P_a,
\eea
where $a$ runs over a finite range.
The simplest type of history is represented by a class operator $C_{\a}$
which is a time-ordered string
of projections
\beq
C_{\a} = P_{a_n} (t_n) \cdots P_{a_1} (t_1).
\label{1.3}
\eeq
Here the projections are in the Heisenberg picture and $ \a $ denotes
the string $ (a_1, \cdots a_n)$.
The class operator Eq.(\ref{1.3}) satisfies the conditions
\beq
\sum_{\a} C_{\a} = 1,
\label{1.4}
\eeq
and also
\beq
\sum_{\a} C^{\dag}_{\a} C_{\a} = 1.
\label{1.5}
\eeq
For a given pure initial state, we may define ``history states'',
\beq
| \a \rangle = C_{\a} | \psi \rangle.
\label{hist}
\eeq
Probabilities are assigned to histories via the formula
\beq
p(\a) =  \langle \a | \a \rangle,
\label{1.6}
\eeq
which is just the Born rule for history states.
These probabilities are clearly positive and normalized
\beq
\sum_{\a} p(\a ) = 1,
\label{1.6b}
\eeq
which follows from Eq.(\ref{1.5}).

As the double slit experiment indicates, the assignment
of probabilities to histories in quantum mechanics is not always
possible. For the $p (\a) $ to be true probabilities
the histories must satisfy certain conditions which, loosely speaking,
ensure that there are no interference effects.
To this end, we introduce the decoherence functional
\beq
D(\a, \a') = \langle \a | \a' \rangle,
\eeq
which may be thought of as a measure of interference between pairs of histories.
It satisfies the conditions
\bea
D(\a, \a') &=& D^* (\a', \a),
\\
\sum_{\a} \sum_{\a'} D(\a, \a') &=& 1,
\label{1.10}
\eea
and note that the probabilities are given by its diagonal elements
\beq
p(\a) = D(\a, \a).
\eeq

The simplest and most important condition normally
imposed is that the probabilities should satisfy the probability sum rules,
that is,
that they are additive for all disjoint pairs of histories.
More precisely, the probability of history $ \a$ or history $\a'$
must be the sum of $p(\a)$ and $p(\a')$. Since this combination
of histories is represented by the class operator $C_{\a} + C_{\a'}$, Eq.(\ref{1.6})
implies that
\beq
p( \a \ {\rm or} \  \a') = p(\a) + p(\a') + 2 \ {\rm Re} D(\a, \a').
\eeq
Hence for the probabilities to satisfy the expected sum rules
we require that
\beq
{\rm Re} D(\a, \a') = 0, \ \ \ \a  \ne \a',
\label{1.13}
\eeq
for all pairs of histories $\a, \a'$. This condition is called
{\it consistency} of histories.
Consistency of histories
ensures that the
probabilities defined by Eq.(\ref{1.6}) satisfy all the conditions
one would expect of a probability for histories. It means that the statements corresponding to the histories may be manipulated according to the rules of Boolean logic.

So far, nothing has been said about measurements. Despite the formal similarities with sequential measurements in quantum measurement theory, in the decoherent histories approach the projections are {\it not} in general regarded as corresponding to actual measurements. Instead, the formalism is used to determine whether, using ordinary logic, we can extrapolate from physically measured quantities to quantities that are not measured. One could, for example, take the final projection to correspond to measured data and then ask to what extent we can deduce anything about the earlier unmeasured alternatives in the history.

A more concrete connection between physical measurements and the alternatives at all times comes into view when we
consider
the stronger condition of {\it decoherence}, which is
\beq
D(\a, \a') = 0, \ \ \ \a \ne \a'.
\eeq
This of interest since physical mechanisms (such as coupling to an environment) which cause Eq.(\ref{1.13}) to be satisfied typically suppress both the real and imaginary off-diagonal parts of the decoherence functional.
This stronger condition is related to the existence of records \cite{GH2,Hal4}. This means that complete information about the entire history of the system is stored somewhere in the system at a fixed moment of time. A good example is a photographic plate showing the track of an elementary particle. If records exist it then means that full information about the histories may be obtained by a single projective measurement.

In more mathematical terms, decoherence clearly means that the history states $|\a \rangle$ are orthogonal. This has the consequence that
one can always find a projection operator $R_{\b}$ whose eigenstates are the history states:
\beq
R_{\b} | \a \rangle = \delta_{ \a \b} | \a \rangle.
\eeq
Summing over $\a$, this means that
\beq
R_{\a} | \psi \rangle = C_{\a} | \psi \rangle.
\eeq
This therefore means that we can add an extra projector $R_{\b}$ at the end of the history which does not disturb decohence and  is
perfectly
correlated with the alternatives at earlier times. That is
\beq
\langle \psi | C_{\a'}^{\dag} R_{\b} C_{\a}  | \psi \rangle
= \delta_{\b \a} \delta_{\b \a'} p (\a).
\eeq
This means that their exists a joint probability distribution $p(\a, \b)$ for the histories $ \a$ and the records $\b$
in which $\a$ and $\b$ are perfectly correlated
\beq
p(\a, \b) = \delta_{\a \b} p(\a) = \delta_{\a \b} p(\b).
\eeq
The probabilities for the histories can then be expressed in terms of a single projection
\beq
p(\a) = \langle \psi | R_{\a} | \psi \rangle.
\eeq
This all has the important consequence that, when there is decoherence, the probabilities for histories can be determined by a single projective measurement at one moment of time.

This feature of the decoherent histories approach is clearly very appealing in relation to measurements of the correlation functions in the LG inequalities, since the measurements have to be done non-invasively, and it is the sequential nature of the measurements that creates the invasiveness. Here, however, we see that a string of projections can, under the condition of decoherence, be equivalent to a single projection.

The record projector is generally non-unique. However, an example which does the job is to construct it from the history states themselves,
\beq
R_{\a} = \frac {  C_{\a} | \psi \rangle \langle \psi | C^\dag_{\a}  } { p(\a) }.
\label{rec}
\eeq
For very simple systems such as the ones considered in this paper, this is in fact the unique choice.

\bibliography{apssamp}

\end{document}